\documentclass[12pt]{article}
\usepackage{fullpage}
\usepackage{graphicx}
\newcommand{\be}{\begin{equation}}
\newcommand{\ee}{\end{equation}}

\newcommand{\pauli}{\mbox{\boldmath $\tau$}}
\newcommand{\bpi}{\mbox{\boldmath $\pi$}}
\newcommand{\I}{{\cal I}}

\newcommand{\news}{\setcounter{equation}{0}}
\def\ben{\begin{equation}}
\def\een{\end{equation}}
\def\bea{\begin{eqnarray}}
\def\eea{\end{eqnarray}}
\begin{document}
\title{
\begin{flushright}\ \vskip -2cm {\normalsize{\em DCPT-08/59}}\end{flushright}
\vskip 2cm Multi-Skyrmions with Vector Mesons}
\author{Paul Sutcliffe\\[10pt]
{\em \normalsize Department of Mathematical Sciences,
Durham University, Durham DH1 3LE, U.K.}\\
{\normalsize Email: p.m.sutcliffe@durham.ac.uk}
}
\date{March 2009}
\maketitle
\begin{abstract}
It is known that including vector mesons stabilizes the size of a 
Skyrmion without the need for a Skyrme term. This paper provides
the first results for static multi-Skyrmions in such a theory.
The rational map ansatz is used to investigate multi-Skyrmions 
in a theory which includes the $\omega$ vector meson and 
has no Skyrme term. Bound states with baryon numbers two, three and four
are found, which have axial, tetrahedral and cubic symmetries, respectively.
The results reveal a qualitative similarity with the standard Skyrme 
model with a Skyrme term and no vector mesons, suggesting that
some features are universal and do not depend on the details of the theory.
Setting the pion decay constant and meson masses to their experimental values
leaves only a single free parameter in the model. Fixing this parameter, 
by equating the energy of the baryon number four Skyrmion to the
$\rm{He}^4$ mass, yields reasonable results for other baryon numbers.
\end{abstract}

\newpage
\section{Introduction}\news\ \
Skyrmions are topological solitons that model baryons within a 
nonlinear theory of pions, arising as a low energy effective theory from QCD
in the limit of a large number of colours \cite{Wi}. 
The standard Skyrme model \cite{Sk} includes only the pion degrees of freedom
and the Lagrangian requires the inclusion of a Skyrme term, which is quartic 
in derivatives. The role of the Skyrme term is to balance the sigma 
model contribution and provide a scale for the
soliton, as required by Derrick's theorem \cite{De}. 
The Skyrme term
has some drawbacks, for example, it makes the theory non-renormalizable
and the classical dynamical field equations have potential instabilities
associated with the loss of hyperbolicity \cite{CB}.

Over twenty years ago it was shown \cite{AN2} that by generalizing 
the nonlinear pion theory to include vector mesons, the size of a Skyrmion
is stabilized without the need for a Skyrme term. The original work
included only the $\omega$ meson but later extensions added other
vector mesons, for example $\rho$ mesons 
\cite{MZ,BKY,HY}. These investigations
produced promising results and revealed some improvements over the standard 
Skyrme model, although all these studies were limited to the sector with 
baryon number one.
 
The single Skyrmion solution is spherically symmetric and therefore it can be 
constructed by solving only ordinary differential equations; though even
these must be solved numerically. However, multi-Skyrmions are not spherically
symmetric and therefore highly nonlinear partial differential equations
in three-dimensional space must be solved to study baryon numbers greater 
than one. The substantial difficulties that need to be overcome mean that 
even today there are still no results available on static multi-Skyrmions in 
theories including vector mesons; though it has been demonstrated that a 
product ansatz allows well-separated Skyrmions to be placed in an
attractive channel \cite{PRV}, and recently a Skyrme crystal has been
investigated \cite{PRV2}.  
 The purpose of the present paper is to 
provide the first results on static multi-Skyrmions in 
theories including vector mesons.
 Details are presented 
for baryon numbers from one to four, but the methods described are also 
applicable to larger baryon numbers.

Substantial progress in both numerical and analytic approaches 
to standard Skyrmions means that multi-Skyrmions are now
fairly well understood in the Skyrme model (for a review see \cite{book}).
The approach taken in this paper
is to apply some of these techniques, in particular the rational map
ansatz \cite{HMS}, to study multi-Skyrmions in a theory without the
Skyrme term but including the $\omega$ vector meson.
Briefly, the rational map ansatz will be applied to provide an
approximation to the Skyrme field and the $\omega$ field will be computed
using an expansion in terms of symmetry adapted spherical harmonics.
 
Bound states with baryon numbers two, three and four
are found, which have axial, tetrahedral and cubic symmetries, respectively.
The results reveal a qualitative similarity with the standard Skyrme 
model, suggesting that some features are universal and do not depend on the 
details of the theory.
Setting the pion decay constant and meson masses to their experimental values
leaves only a single free parameter in the model. Fixing this parameter, 
by equating the energy of the baryon number four Skyrmion to the
$\rm{He}^4$ mass, yields reasonable results for other baryon numbers.

Finally, recent developments in AdS/QCD have 
led to renewed interest in baryons as Skyrmions \cite{SS}.
In particular, these studies point to the importance of the inclusion of 
vector mesons and provide a string theory motivation for old ideas of 
vector meson dominance.

\section{Skyrmions and the $\omega$ meson}\news\ \
The Skyrme field $U$ takes values in $SU(2)$ and satisfies the boundary
condition that $U\rightarrow 1$ as $|{\bf x}|\rightarrow \infty.$
It is related to the triplet of pion fields $\bpi$ through the formula
\be
U=\sigma +i\bpi\cdot\pauli,
\ee
where $\pauli$ denotes the triplet of Pauli matrices and 
 $\sigma^2+\bpi\cdot\bpi=1.$

Topological solitons arise because there is a conserved topological current  
\be
B^\mu=\frac{1}{24\pi^2}\epsilon^{\mu\nu\alpha\beta}
Tr(\partial_\nu U\,U^\dagger\,\partial_\alpha U\,U^\dagger\,
\partial_\beta U\,U^\dagger),
\label{current}
\ee
and the associated integer topological charge $B=\int B^0\, d^3x$
is the soliton number and is identified with baryon number.

The theory for the Skyrme field coupled to the $\omega$ vector meson 
is given by the Lagrangian density \cite{AN2}
\be
{\cal L}=\frac{F_\pi^2}{16}\mbox{Tr}(\partial_\mu U\partial^\mu U^\dagger)
+\frac{F_\pi^2 m_\pi^2}{8}\mbox{Tr}(U-1)
-\frac{1}{4}(\partial_\mu\omega_\nu-\partial_\nu\omega_\mu)
(\partial^\mu\omega^\nu-\partial^\nu\omega^\mu)
+\frac{m_\omega^2}{2}\omega_\mu\omega^\mu+\beta\omega_\mu B^\mu,
\label{lagparam}
\ee
where $F_\pi=186\,\rm{MeV}$ is the pion decay constant, $m_\pi=138\,\rm{MeV}$ 
is the pion mass and $m_\omega=782\,\rm{MeV}$ is the $\omega$ mass.
The constant $\beta$ can be related to the $\omega\rightarrow 3\pi$
decay rate, however, this is enhanced by the resonance 
$\omega\rightarrow \rho +\pi$ which is not included in the current theory.
Therefore the experimental data only provides an upper bound on $\beta,$
which is found  to be $\beta\le 25.4 $ \cite{AN2}.

It is convenient to remove various constants in the above Lagrangian density
by rescaling the $\omega$ field as $\omega\mapsto \omega F_\pi$
and using energy and length units of $F_\pi^2/m_\omega$ and $1/m_\omega$
respectively. The Lagrangian density then becomes
\be
{\cal L}=\frac{1}{16}\mbox{Tr}(\partial_\mu U\partial^\mu U^\dagger)
+\frac{M^2}{8}\mbox{Tr}(U-1)
-\frac{1}{4}(\partial_\mu\omega_\nu-\partial_\nu\omega_\mu)
(\partial^\mu\omega^\nu-\partial^\nu\omega^\mu)
+\frac{1}{2}\omega_\mu\omega^\mu+g\omega_\mu B^\mu,
\label{lag}
\ee
where $M=m_\pi/m_\omega=0.176$ and $g=\beta m_\omega /F_\pi.$

Note that the theory (\ref{lag}) does not contain a Skyrme term
and has only one free parameter, $g,$ once the meson masses are
fixed to their experimental values.

Several values for $g$ have been studied and the results are qualitatively 
similar in all cases. Results will be presented for the two values 
$g=96.7$ and $g=34.7,$ which are both consistent with the
above upper bound for $\beta.$ The first value, which is adopted from
now on until further notice, is taken from \cite{AN2}, and is obtained by
allowing both $g$ and $F_\pi$ to be free parameters whose values are 
obtained by fitting the masses of the nucleon and delta resonance to the
energies of a spinning $B=1$ Skyrmion. This approach results in a value
of $F_\pi$ that is  $124\,\rm{MeV}$ and therefore a little less than the
experimental result.
 The second value of $g$ is calculated in a later section 
by setting $F_\pi$ to the experimental value and fitting the energy of
the $B=4$ Skyrmion to the $\rm{He}^4$ mass.

Only static fields are considered in this paper, therefore the spatial
components of the topological current vanish $B^i=0.$ 
As the topological current provides the source term for the field $\omega_\mu$ 
then its spatial components can be set to zero, $\omega_i=0.$ 
For ease of notation, in the following we drop the subscript on the
temporal component and write $\omega\equiv \omega_0.$
  
Taking into account the above comments, the static energy associated with
the Lagrangian density (\ref{lag}) is given by
\be
E=\int\left(
\frac{1}{16}\mbox{Tr}(\partial_i U\partial_i U^\dagger)
+\frac{M^2}{8}\mbox{Tr}(1-U)
-\frac{1}{2}\partial_i\omega\partial_i\omega
-\frac{1}{2}\omega^2-g\omega B^0\right)
 d^3x.
\label{energy}
\ee
Note the negative signs associated with the energy in the $\omega$ field,
as it is a temporal component.

Variation of the above energy with respect to $\omega$ yields the 
field equation 
\be
\partial_i\partial_i\omega-\omega =gB^0,
\label{oeom}
\ee
which is a linear equation for $\omega$ with a source term proportional 
to the topological charge density. The boundary condition is that
$\omega$ vanishes at spatial infinity.

The problem at hand is therefore one of constrained energy minimization.
The Skyrme field $U$ is obtained as the field configuration, with a given
topological charge $B,$ that minimizes the energy (\ref{energy}),
with $\omega$ determined uniquely by $U$ as the solution of equation
(\ref{oeom}). In the following section an approximation technique is described
and implemented to solve this problem.  

Using equation (\ref{oeom}), together with an integration by parts, the energy 
(\ref{energy}) can also be written as
\be
E=\int\left(
\frac{1}{16}\mbox{Tr}(\partial_i U\partial_i U^\dagger)
+\frac{M^2}{8}\mbox{Tr}(1-U)-\frac{1}{2}g\omega B^0\right)
 d^3x,
\label{energy2}
\ee
which will be useful later. Alternatively, the same procedure can be
used to remove the last term in expression (\ref{energy}) in exchange
for changing the two minus signs to plus signs in front of the 
$\omega$ dependent contributions. 
This form makes it transparent that the additional contribution
to the energy from the inclusion of the $\omega$ field is non-negative.

\section{Constructing multi-Skyrmions}\news\ \
Extensive numerical computations of the full nonlinear field equations
of the standard Skyrme model, with massless pions, have determined the minimal 
energy Skyrmions with baryon numbers up to $B=22$ \cite{BS3}.
These numerical results can be reproduced with a surprising accuracy
using an approximation employing rational maps between Riemann spheres,
known as the rational map ansatz \cite{HMS}. This ansatz is briefly reviewed
below.

In terms of the standard spherical polar coordinates $r,\theta,\phi,$ 
introduce the Riemann sphere coordinate $z=e^{i\phi}\tan(\theta/2),$ given by
stereographic projection of the two-sphere. 
Let $R(z)$ be a degree $B$ rational map between Riemann spheres,
that is, $R=p/q$ where $p$ and $q$ are polynomials in $z$ such that
$\max[\mbox{deg}(p),\mbox{deg}(q)]=B$,  and $p$ and $q$ have no common
factors.  Given a rational map $R(z)$ the ansatz for the Skyrme field is
\be  U(r,z)=\exp\bigg[\frac{if(r)}{1+\vert R\vert^2} \pmatrix{1-\vert
R\vert^2& 2\bar R\cr 2R & \vert R\vert^2-1\cr}\bigg]\,,
\label{rma}
\ee where $f(r)$ is a real profile function which satisfies the boundary
conditions $f(0)=\pi$ and $f(\infty)=0.$ 

Applying the ansatz (\ref{rma}) to the baryon density (\ref{current})
produces the expression
\be
B^0=-\frac{f'}{2\pi^2}\left(\frac{\sin f}{r}
\frac{1+|z|^2}{1+|R|^2}\left|\frac{dR}{dz}\right|\right)^2,
\label{bden}
\ee
from which it is simple to verify that the topological charge is indeed
$B,$ as a consequence of the boundary conditions on $f(r)$ and the 
fact that $R(z)$ has degree $B.$

For the particular choice $R=z,$ the ansatz (\ref{rma}) reduces to
the standard hedgehog ansatz for a spherically symmetric $B=1$ Skyrmion.
For $B>1$ this ansatz does not provide any exact solutions,
but for suitable choices of rational maps and
profile functions, it provides excellent approximations to the true
multi-Skyrmion solutions of the standard Skyrme model.
In particular, the ansatz reproduces the correct symmetries of the
true solutions, and the energy of the ansatz typically only overestimates the
true energy of a multi-Skyrmion by the order of a percent.

Substituting the ansatz (\ref{rma}) into the energy of the 
standard Skyrme model leads to only one contribution which is sensitive to
the properties of the rational map beyond its degree. This contribution
originates entirely from the Skyrme term and involves the coefficient  
\be \I=\frac{1}{4\pi}\int \bigg(
\frac{1+\vert z\vert^2}{1+\vert R\vert^2}
\bigg\vert\frac{dR}{dz}\bigg\vert\bigg)^4 \frac{2i \  dz  d\bar z
}{(1+\vert z\vert^2)^2}\,.
\label{i}
\ee 
Thus, within the rational map ansatz, the problem of finding
the minimal energy Skyrmion in the standard Skyrme model 
reduces to the simpler problem of
calculating the rational map which minimizes the function
$\I$. Given the minimizing rational map, or more precisely the
associated value of $\I,$ the profile function can then be determined
by minimizing an energy functional for $f(r).$

The $\I$ minimizing rational map for $B=1$ is the spherically symmetric 
map $R=z$ and for $B=2$ it is the axially symmetric map $R=z^2,$
in agreement with the fact that the minimal energy $B=2$ Skyrmion
is axially symmetric \cite{KS,Ma3,Ve}. 

For any $B>1$ the map $R=z^B$ is axially symmetric, but it is not the
$\I$ minimizing map for $B>2.$ For $B=3$ and $B=4$ the $\I$ minimizing
maps are the unique maps with tetrahedral and cubic symmetry, 
respectively, and are given by
\cite{HMS}
\be
R=\frac{z^3-\sqrt{3}iz}{\sqrt{3}iz^2-1},
\quad\quad\quad
R=\frac{z^4+2\sqrt{3}iz^2+1}{z^4-2\sqrt{3}iz^2+1}.
\label{maps}
\ee
These symmetries again agree with those of the numerically computed
Skyrmions in the standard Skyrme model \cite{BTC}.

In the vector meson model there is no Skyrme term and hence it
appears that the coefficient $\I$ is not relevant.
However, notice that if the Laplacian term is neglected
in the $\omega$ field equation (\ref{oeom}) then $\omega$
is simply equal to a negative constant times the baryon density.
Substituting this approximation into the energy (\ref{energy2})
gives that the interaction term is a positive constant times
the square of the baryon density. With the rational map expression
for the baryon density (\ref{bden}) the angular part of this energy
is precisely $\I,$ and therefore this term arises in the vector meson
model as a leading order contribution in a derivative expansion.
Alternatively, a formal solution for $\omega$ can be written in
terms of the Green's function for the massive Klein-Gordon equation.
If this representation is applied to the interaction term in the energy 
then it becomes a non-local expression involving two factors of the
baryon density. The angular part of this is therefore a non-local
version of $\I,$ which reproduces $\I$ exactly in the limit in which the
Green's function is replaced by a delta function.   

The above arguments suggest that the $\I$-minimizing maps are also
the appropriate maps for the vector meson model, and indeed it  
 will be shown that the maps (\ref{maps}) yield low energy 
bound states, in contrast to the maps $R=z^3$
and $R=z^4,$ for example. This is strong evidence that in the vector
meson model the minimal energy Skyrmions for $B=3$ and $B=4$ have the
same Platonic symmetries as in the standard Skyrme model.    
Additional evidence is provided by studying a (2+1)-dimensional
analogue of this problem, where numerical solutions of the exact static
field equations reveal an amazing similarity between solitons 
in the Baby Skyrme model and its vector meson version \cite{FS}.

The approach applied below involves exploiting the symmetry of the Skyrme
field to solve the field equation for $\omega$ by using an expansion
in terms of symmetry adapted spherical harmonics. This aspect is 
reviewed in the following.

Given a subgroup $G\subset SO(3)$ of spatial rotations, introduce
the symmetric harmonics $K_l(\theta,\phi),$ as a set of real 
orthonormal functions, 
each of which is a linear combination of spherical harmonics  
$Y_{lm}(\theta,\phi),$  that is, 
\be
K_l=\sum_{m=-l}^l \alpha_{lm}Y_{lm},
\label{defk}
\ee
where the coefficients $\alpha_{lm}$ are chosen so that 
each function $K_l$ is invariant under the group $G.$
Strictly speaking, the functions $K_l$ should carry
an additional index, since there may be more than 
one symmetric harmonic with a given value of $l.$ 
However, for the specific calculations and symmetries discussed below,
 only the values $0\le l\le 10\,$
will be required, and there is a unique symmetric harmonic for each $l$
in this range; therefore for simplicity the additional index will be 
suppressed.

As an example, if $G=SO(2),$  then the axially symmetric harmonics are 
simply $K_l=Y_{l0}.$ 

The angular part of the rational map generated baryon density (\ref{bden}) 
is given by
\be
b(\theta,\phi)=\frac{1}{4\pi}\left(\frac{1+|z|^2}{1+|R|^2}\left|\frac{dR}{dz}
\right|\right)^2,
\label{b}
\ee
where the normalization is such that the integral of $b$ over the two-sphere
is equal to $B.$

For a $G$-symmetric rational map the angular baryon density $b$ 
can be expanded in terms of symmetric harmonics as
\be
b=\sum_l b_l K_l,
\label{expandb}
\ee
where $b_0=B/(2\sqrt{\pi})$ from the chosen normalization.
Similarly, $\omega$ can also be expanded in terms of symmetric harmonics
as 
\be
\omega=\sum_l h_l(r)K_l,
\label{expando}
\ee
where $h_l(r)$ are real profile functions which depend only on the radial
coordinate $r.$ The boundary conditions are that $h_l(\infty)=0$ and
$h_l(0)=0$ for $l>0$ with $h_0'(0)=0.$  

The baryon density expression (\ref{bden}) together with the expansions
(\ref{expandb}) and (\ref{expando}) reduces the $\omega$ field equation 
(\ref{oeom}) to a set of profile function equations
\be
h_l''+\frac{2}{r}h_l'-\left(1+\frac{l(l+1)}{r^2}\right)h_l'
=-\frac{2g}{\pi}\frac{f'\sin^2f}{r^2}b_l.
\label{profileh}
\ee
Substituting the rational map ansatz (\ref{rma}) and the expansions 
(\ref{expandb}) and (\ref{expando}) into the energy (\ref{energy})
and performing the angular integration results in
\be
E=E_\omega+\frac{\pi}{2}\int_0^\infty\{
r^2f'^2+2B\sin^2f+2M^2r^2(1-\cos f)
+\frac{4g}{\pi^2} f'\sin^2f \sum_l b_lh_l\}dr,
\label{energy3}
\ee
where $E_\omega$ denotes a contribution to the energy 
which has no explicit $f$ dependence.

The equation for $f$ obtained from the variation of the energy 
(\ref{energy3}) is 
\be
f''+\frac{2}{r}f'-\frac{B}{r^2}\sin 2f-M^2\sin f
+\frac{2g}{\pi^2} \frac{\sin^2f}{r^2} \sum_l b_lh_l'=0.
\label{profilef}
\ee

For functions $h_l$ and $f$ which satisfy equations
(\ref{profileh}) and (\ref{profilef}) the energy expression
(\ref{energy2}) can be used to rewrite the energy as 
\be
E=\frac{\pi}{2}\int_0^\infty\{
r^2f'^2+2B\sin^2f+2M^2r^2(1-\cos f)
+\frac{2g}{\pi^2} f'\sin^2f \sum_l b_lh_l\}dr,
\label{energy4}
\ee
which is similar to (\ref{energy3}) except that
there is no longer a term which is independent of $f$ and
the coefficient in the final term has been halved.

Given a $G$-symmetric rational map $R(z)$ the angular baryon density
(\ref{b}) can be calculated and the coefficients $b_l$ in the expansion
(\ref{expandb}) computed. The profile functions $h_l$ and $f$ can then
be found by numerically solving the equations (\ref{profileh}) and
(\ref{profilef}), using a heat flow algorithm.
Finally, these profile functions are used to determine the
 energy from the expression (\ref{energy4}). As mentioned earlier, it is 
found that truncating the expansions at $l=10$ is sufficient to produce
results of the required accuracy.

\begin{figure}
\begin{center}
\includegraphics[width=11cm]{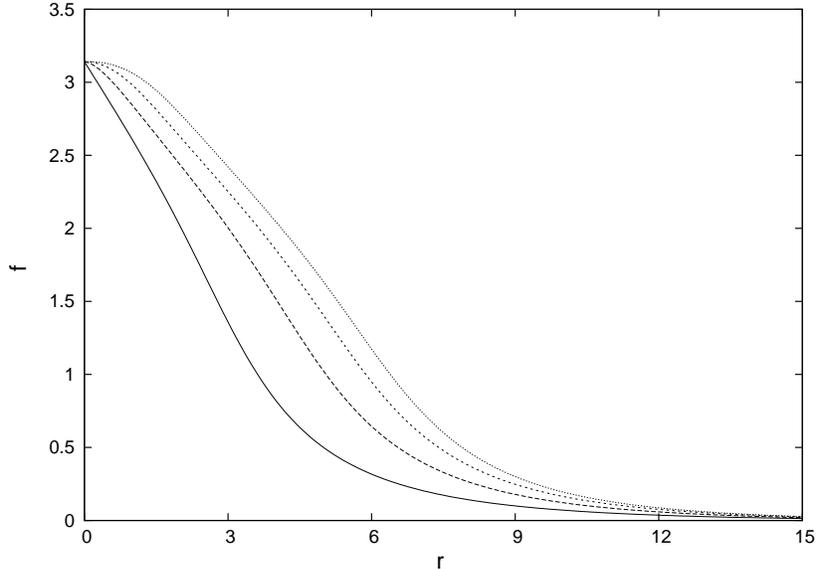}
\caption{Profile functions $f(r)$ for $B=1,2,3,4.$ The 
curves shift to the right with increasing $B$.
}
\label{fig-f}
\end{center}
\end{figure}

\begin{figure}
\begin{center}
\includegraphics[width=11cm]{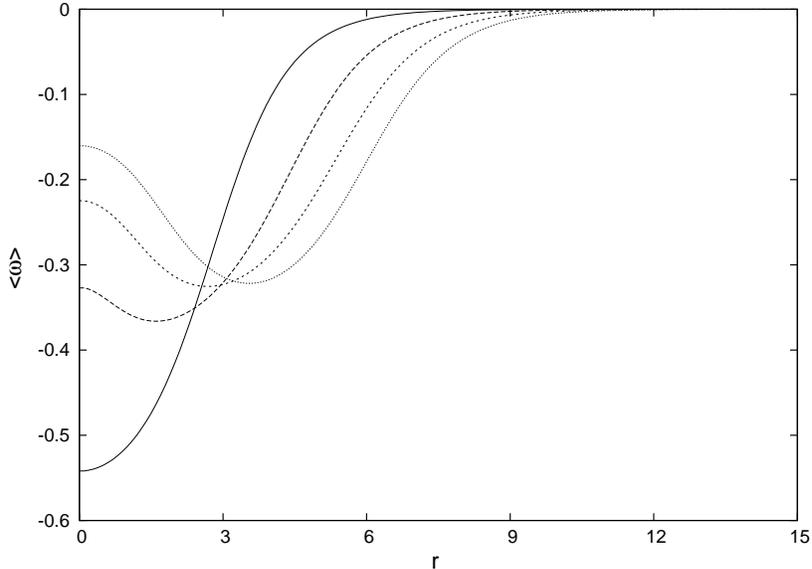}
\caption{The spherical average $<\omega>$ as a function of the radius $r$,
 for $B=1,2,3,4.$ The curves shift up and to the right with increasing $B$.
}
\label{fig-o}
\end{center}
\end{figure}

The simplest example is the spherical $B=1$ Skyrmion with rational
map $R=z.$ There is only one spherically symmetric harmonic
$K_0=Y_{00}=1/(2\sqrt{\pi})$ and hence only the profile functions $f$ and
$h_0$ need to be calculated. In this case the rational map ansatz is
exact and the above procedure reproduces the earlier result \cite{AN2}.
For $g=96.7$ the energy is found to be $E_1=44.04$ and the 
associated profile function
$f$ is plotted as the solid curve in Figure~\ref{fig-f}. 
The spherically symmetric field $\omega=h_0K_0$ is presented as the 
solid curve in Figure~\ref{fig-o}.

Next consider the axially symmetric $B=2$ map $R=z^2.$
 As mentioned earlier, the axially symmetric harmonics are 
$K_l=Y_{l0}.$ However, the baryon density has an additional symmetry
under rotations by $180^\circ$ around an axis orthogonal to
the $SO(2)$ symmetry axis, which implies
that only symmetric harmonics with even $l$ are needed.  
A computation of the angular baryon density produces the expansion
coefficients
\be
b=0.564K_0-0.363K_2+0.107K_4-0.026K_6+0.006K_8-0.001K_{10}+\ldots
\ee 

A numerical solution of the profile function equations (for $l\le 10$) 
results in the energy 
$E_2=87.44<88.08=2E_1,$ and hence this is a state which is bound 
against the break-up
into two single Skyrmions. Note that the binding energy is quite small,
being less than 1\%. However, the rational map ansatz is exact
for a single Skyrmion and is only an approximation for multi-Skyrmions,
therefore any computation of binding energies which compares to single
Skyrmions will be a slight underestimate. 

\begin{figure}
\begin{center}
\includegraphics[width=10cm]{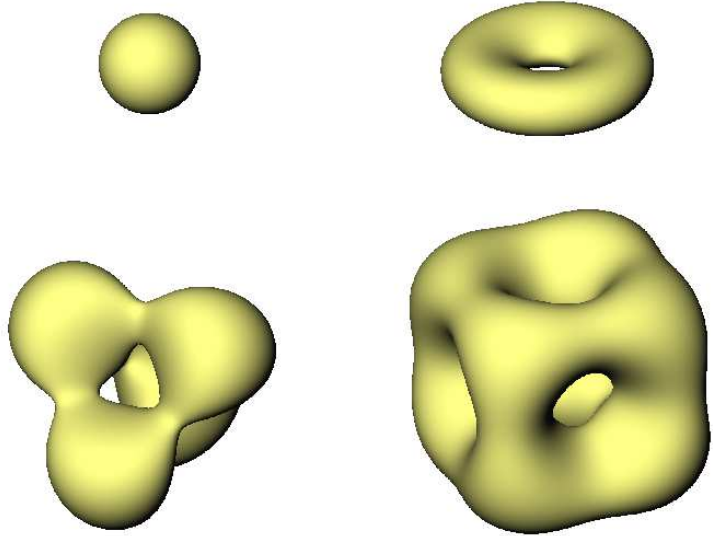}
\caption{Isosurfaces of $\omega$ for $B=1,2,3,4$ (to scale). 
The isosurface values are given by $B=1$, $\omega=-0.44$;\
$B=2$, $\omega=-0.40$;\ $B=3$, $\omega=-0.30$;\ $B=4$, $\omega=-0.28.$
}
\label{fig-all4}
\end{center}
\end{figure}

The profile function $f(r)$ is displayed in Figure~\ref{fig-f}
(the curves shift to the right with increasing $B$).
Although the $\omega$ field now has a non-trivial angular dependence, 
it is useful to plot the spherically averaged value 
\be
<\omega>=\frac{1}{4\pi}\int\omega \sin\theta\,d\theta\,d\phi,
\ee
which is displayed in Figure~\ref{fig-o}
 (the curves shift up and to the right with increasing $B$).
This reveals that although the $\omega$ field is again large and
negative at the origin it takes its most negative value at a positive
radius. This is to be expected because the source term for $\omega$ is
the baryon density and this has a toroidal distribution.
An isosurface plot of $\omega$ is displayed as the top right image 
in Figure~\ref{fig-all4}, and the toroidal distribution is evident.
Although isosurface plots of $\omega$ resemble the associated baryon density
surfaces there are some differences, for example, the baryon density
vanishes at the origin for $B>1$ whereas, as seen above, the $\omega$
field at the origin is substantial.  

Axially symmetric fields for larger baryon numbers can be studied
by applying the same procedure as above to the rational map $R=z^B.$
However, it is found that all axially symmetric fields of this type
are unbound against the break-up into $B$ single Skyrmions. For example,
for $B=3$ the energy is calculated to be 
$E_3^{{\rm axial}}=138.24>132.12=3E_1.$ 
Of course, since the rational map approximation has been used to 
compute this energy it is expected
that the true energy of the axial $B=3$ field is slightly less than the
above value, but the correction is unlikely to be large enough to
yield a bound state. This is strong evidence that the $B=3$ Skyrmion
is not axially symmetric, in agreement with the usual Skyrme model.
As mentioned earlier, with a Skyrme term and no vector mesons
the minimal energy Skyrmion with $B=3$ has tetrahedral symmetry
and is described by the first rational map in (\ref{maps}). 
In the following it is shown that
in the $\omega$ meson theory a tetrahedral bound state also exists
with baryon number three, which suggests that it is the
minimal energy $B=3$ Skyrmion.

To study Platonic Skyrmions the symmetric harmonics $K_l$ are
required, where $G$ is one of the Platonic symmetry groups.
The construction of Platonic harmonics using invariant generating polynomials
was introduced by Bethe and collaborators \cite{Be,VB}, and this is
briefly reviewed below for the tetrahedral case.
 
The ring of tetrahedrally invariant homogeneous polynomials is generated by 
three polynomials of degrees two, three and four,
\be
p_2=x_1^2+x_2^2+x_3^2, \quad\quad
p_3=x_1x_2x_3, \quad\quad
p_4=x_1^4+x_2^4+x_3^4. 
\label{genpoly}
\ee
To obtain a tetrahedral harmonic $K_l,$ the first step is to determine
the most general degree $l$ homogeneous polynomial, $k_l$, 
that can be constructed as a linear combination of products of the 
polynomials (\ref{genpoly}). The coefficients in the polynomial $k_l$
are then fixed (up to an irrelevant overall factor) by requiring that
$k_l$ solves Laplace's equation. 

Note that for $l=2,3,5$ there is only one possible contribution to
 $k_l,$ namely $p_2,\, p_3$ and $p_2p_3$ respectively. Of these three
polynomials only $p_3$ satisfies Laplace's equation, hence there are
no tetrahedral harmonics with $l=2$ or $l=5,$ and obviously none with $l=1.$

Given the tetrahedrally invariant polynomial $k_l$ then $k_l/r^l$ is
a function of only the angular coordinates $\theta,\phi$ and, with
a suitable normalization, is the required tetrahedral harmonic $K_l,$
which can be decomposed into spherical harmonics $Y_{lm}.$ 
In addition to the above construction, Platonic harmonics may also be
determined using a projector technique \cite{Wo}, but in either case the
results are
\bea
& &
K_0=Y_{00},\quad
K_3=\frac{i}{\sqrt{2}}(\bar Y_{32}-Y_{3-2}),\quad
K_4=\frac{1}{2}\sqrt{\frac{5}{6}}\left(
Y_{44}+Y_{4-4}+\sqrt{\frac{14}{5}}Y_{40}
\right),\quad \nonumber \\
& &
K_6=\frac{\sqrt{7}}{4}\left(
Y_{64}+Y_{6-4}-\sqrt{\frac{2}{7}}Y_{60}
\right),\quad
K_7=i\frac{\sqrt{33}}{12}\left(
\bar Y_{76}-Y_{7-6}+\sqrt{\frac{13}{11}}(\bar Y_{72}-Y_{7-2})
\right),\quad \nonumber \\
& &
K_8=\frac{1}{24}\sqrt{\frac{195}{2}}\left(
Y_{88}+Y_{8-8}+2\sqrt{\frac{7}{65}}(Y_{84}+Y_{8-4})
+3\sqrt{\frac{22}{65}}Y_{80}
\right),\quad  \label{tetharm}\\
& &
K_9=\frac{i}{4}\sqrt{\frac{13}{2}}
\left(
\bar Y_{96}-Y_{9-6}-\sqrt{\frac{3}{13}}(\bar Y_{92}-Y_{9-2})
\right),\quad \nonumber \\
& &
K_{10}=\frac{\sqrt{561}}{48}\left(
Y_{10,8}+Y_{10,-8}+\frac{6}{\sqrt{51}}(Y_{10,4}+Y_{10,-4})
-\sqrt{\frac{130}{187}}Y_{10,0}
\right).\nonumber
\eea 

Substituting the first rational map in (\ref{maps}) into the
angular baryon density (\ref{b}) and expanding in terms of
tetrahedral harmonics produces
\be
b=0.846K_0-0.585K_3-0.070K_4-0.127K_6+0.026K_7
+0.002K_8+0.022K_9
+0.006K_{10}+\ldots
\ee 
Using the above coefficients 
the associated profile functions can be computed
($f$ is presented in Figure~\ref{fig-f} and the spherical
average $<\omega>$ is plotted in Figure~\ref{fig-o}) and the energy
found to be $E_3=126.63<131.48=E_2+E_1,$ and therefore a bound state.
A tetrahedrally symmetric $\omega$ isosurface is displayed as the 
lower left image in Figure~\ref{fig-all4}.

Turning attention to $B=4,$ the required rational map is the second map
in (\ref{maps}) and has cubic symmetry.
The cubically symmetric harmonics are, of course, a subset of those
with tetrahedral symmetry. The tetrahedral polynomials $p_2$ and $p_4$ 
in (\ref{genpoly}) are also invariant under the cubic group, but $p_3$
is not. However, $p_6\equiv p_3^2$ is invariant under the cubic group,
and indeed $p_2,p_4,p_6$ are the generating polynomials. This implies that
the required cubic harmonics are the tetrahedral harmonics
$K_l$ (\ref{tetharm}) with even $l.$

The expansion coefficients of the $B=4$ angular baryon density are given by
\be
b=1.128K_0-0.601K_4-0.050K_6+0.076K_8+0.009K_{10}+\ldots
\ee 
and the results of computing the profile functions are again displayed
in Figure~\ref{fig-f} and Figure~\ref{fig-o}. 
The energy is calculated to be $E_4=159.67,$ and this confirms that 
this configuration is bound against the break-up into all possible 
lower charge clusters. A cubically symmetric $\omega$ isosurface is 
displayed as the lower right image in Figure~\ref{fig-all4}.

The energy results for this coupling, which recall is $g=96.7$,
are summarized in the second and third columns of Table~\ref{tab-energy},
where the energy is also presented as a ratio to that of a single Skyrmion
(this is convenient for the bound state comparison).

\begin{table}[ht]
\centering
\begin{tabular}{|c|c|c|c|c|c|c|}
\hline
& \multicolumn{2}{|c|}{$g=96.7$} &  \multicolumn{4}{|c|}{$g=34.7$} \\
\hline
$B$ & $E_B$ & $E_B/E_1$ & $E_B$ & $E_B/E_1$ & $E_B$ in \rm{MeV}
& Experiment \\
\hline
1 & 44.04 & 1.000 & 22.53 & 1.000 & 996 & 939\\
2 & 87.44 & 1.985 & 45.20 & 2.006 & 1999 & 1876\\
3 & 126.63 & 2.875 & 65.88 & 2.924 & 2913 & 2809\\
4 & 159.67 & 3.626 & 84.28 & 3.741 & 3727 & 3727\\
\hline
\end{tabular}
\caption{The energy,  $E_B,$ of the charge $B$ Skyrmion 
and the ratio $E_B/E_1,$ of the energy to that of a single Skyrmion,
for two different values of the coupling $g.$
For the second value of the coupling the energy is also given in \rm{MeV}
and the corresponding experimental value is listed for comparison.}
 \label{tab-energy}
\end{table}

A range of values for the coupling $g$ have been investigated
and the results are qualitatively similar.
As discussed earlier, the value $g=96.7$ is
taken from \cite{AN2}, which involves fitting properties of the
single Skyrmion to the nucleon and delta, whilst treating $F_\pi$ 
as a free parameter. However, it is known that this method of
parameter fitting can be problematic, in any Skyrme model, because of
difficulties associated with the rigid rotor approximation \cite{BKS}. 
Access to the properties of multi-Skyrmions allows the following 
alternative procedure to be applied. 

All the physical
parameters, meson masses $m_\pi,$ $m_\omega,$ and the 
pion decay constant $F_\pi,$ are set to the experimental values
listed earlier, leaving only the single parameter $g$ to be
determined. A sensible way to fit this parameter is to match
the energy of the $B=4$ Skyrmion to the $\rm{He}^4$ mass, which
is 3727\,\rm{MeV}. The reason this is a good approach is that the
ground state of $\rm{He}^4$ has zero spin and isospin, therefore there
are no issues to address regarding the inclusion of quantum energies
associated with spin and isospin. Adopting this approach yields the value
$g=34.7$ and the associated Skyrmion energies, plus the 
experimental masses for comparison, are listed in columns
four to seven of Table~\ref{tab-energy}.

One point to note is that reducing $g$ leads to a reduction
in the relative binding energies. 
In particular, the data presented in Table~\ref{tab-energy}
reveals that for this new value of the coupling $E_2=2E_1\times1.003,$
and therefore it appears that the $B=2$ Skyrmion is not a bound state.
However, recall that the rational map ansatz is only an approximation
for $B>1$ and the true energy is expected to be anything up to
the order of a percent lower than the approximate value. 
As the value of $E_2$ is so close
to that of $2E_1$ then this correction is expected to produce a $B=2$ 
bound state, with a small binding energy. In fact, experimentally
the deuteron binding energy is around 2\,\rm{MeV}, which is only about
0.1\% of the deuteron mass. Therefore, if a reasonable comparison
with experimental data is to be achieved, then even a tiny overestimate
of the $B=2$ energy would indeed be expected to result in an apparently
unbound state.
  
A comparison of the last two columns in Table~\ref{tab-energy} shows
that all the calculated energies are within 7\% of the experimental
values, which is very reasonable given that the theory
has only a single free parameter $g.$
This accuracy is comparable to that in the standard Skyrme model, when
the pion decay constant is treated as a free parameter whose value
emerges as considerably lower than the experimental one.

The Skyrmion energies presented are the classical energies and 
do not include quantum contributions associated with spin and isospin.
These quantum energies are traditionally added via a zero mode
quantization involving the calculation of inertia tensors 
from the classical solution \cite{ANW}. Expressions for the 
required inertia tensors have been calculated in terms of rational
maps and profile functions \cite{Ko,MMW} and therefore the results
in this paper could be used to calculate these quantum corrections
to the presented classical energies.

\section{Conclusion}\news\ \
Multi-Skyrmions have been investigated in a theory without a Skyrme term
but including the $\omega$ vector meson to provide a scale for the Skyrmion.
The approach employed here involved using 
the rational map ansatz and therefore an obvious
avenue for future research is to perform full field simulations of this
theory, to test the accuracy of the approximations used. 
In particular, in the standard Skyrme model it is known that 
for baryon numbers above seven, there is an important qualitative 
difference between multi-Skyrmions in a theory with massive or
massless pions \cite{BS10,BMS}. It would be of interest to know
if a similar result holds for the theory with vector mesons
and no Skyrme term.

Extending the current analysis to a theory including other vector mesons,
starting with the $\rho$ meson along the lines of \cite{MZ}, is clearly
of importance. Furthermore, the emergence of a Skyrme model with
vector mesons from AdS/QCD \cite{SS} suggests that a Skyrme term
should also be included, even if it is not required to provide a
scale for the Skyrmion. The results presented in this paper suggest
that the vector meson and Skyrme term theories have very similar properties,
and therefore one might expect that a theory which includes both
contributions will also have similar universal features.

\section*{Acknowledgements}
Many thanks to Wojtek Zakrzewski for useful discussions.
I thank the STFC for support under the rolling grant ST/G000433/1.


\begin{thebibliography}{100}

\bibitem{Wi} E. Witten, 
\textit{Nucl. Phys.} \textbf{B223}, 422 (1983);
{\it ibid} \textbf{B223}, 433 (1983).

\bibitem{Sk} T.~H.~R. Skyrme, 
\textit{Proc. R. Soc. Lond.} \textbf{A260}, 127 (1961).

\bibitem{De} G.~H. Derrick, 
\textit{J. Math. Phys.} \textbf{5}, 1252 (1964).

\bibitem{CB} W.~Y. Crutchfield and J.~B. Bell, 
\textit{J. Comp. Phys.} \textbf{110}, 234 (1994).

\bibitem{AN2} G.~S. Adkins and C.~R. Nappi, 
\textit{Phys. Lett.} \textbf{B137}, 251 (1984).

\bibitem{MZ} U.~G. Meissner and I. Zahed,
\textit{Phys. Rev. Lett.} \textbf{56}, 1035 (1986).

\bibitem{BKY} M. Bando, T. Kugo and K. Yamawaki,
\textit{Phys. Reports} \textbf{164}, 217 (1988). 

\bibitem{HY} M. Harada and K. Yamawaki,
\textit{Phys. Reports} \textbf{381}, 1 (2003).

\bibitem{PRV} B.-Y. Park, M. Rho and V. Vento, 
\textit{Nucl. Phys.} \textbf{A736}, 129 (2004).

\bibitem{PRV2} B.-Y. Park, M. Rho and V. Vento, 
\textit{Nucl. Phys.} \textbf{A807}, 28 (2008).

\bibitem{book} N.~S. Manton and P.~M. Sutcliffe,
{\em Topological Solitons}, Cambridge University Press (2004).

\bibitem{HMS} C.~J. Houghton, N.~S. Manton and P.~M. Sutcliffe,
\textit{Nucl. Phys.} \textbf{B510}, 507 (1998).

\bibitem{SS} T. Sakai and S. Sugimoto,
\textit{Prog. Theor. Phys.} \textbf{113}, 843 (2005).

\bibitem{BS3} R.~A. Battye and P.~M. Sutcliffe, 
\textit{Phys. Rev. Lett.} \textbf{79}, 363 (1997);
\textit{Phys. Rev. Lett.} \textbf{86}, 3989 (2001);
\textit{Rev. Math. Phys.} \textbf{14}, 29 (2002). 

\bibitem{KS} V.~B. Kopeliovich and B.~E. Stern,
\textit{JETP Lett.} \textbf{45}, 203 (1987).

\bibitem{Ma3} N.~S. Manton, 
\textit{Phys. Lett.} \textbf{B192}, 177 (1987).

\bibitem{Ve} J.~J.~M. Verbaarschot, 
\textit{Phys. Lett.} \textbf{B195}, 235 (1987).

\bibitem{BTC} E. Braaten, S. Townsend and L. Carson,
\textit{Phys. Lett.} \textbf{B235}, 147 (1990).

\bibitem{FS} D. Foster and P.~M. Sutcliffe, 
{\em Baby Skyrmions stabilized by vector mesons}, arXiv:0901.3622
(2009). 

\bibitem{Be} H. Bethe,
\textit{Ann. d. Physik.} \textbf{395}, 133 (1929).

\bibitem{VB} F.~C. Von der Lage and H.~A. Bethe,
\textit{Phys. Rev.} \textbf{71}, 612 (1947).

\bibitem{Wo} P.~E.~S. Wormer,
\textit{Mol. Phys.} \textbf{99}, 1973 (2001).

\bibitem{BKS} R.~A. Battye, S. Krusch and P.~M. Sutcliffe,
\textit{Phys. Lett. } \textbf{B626}, 120 (2005).

\bibitem{ANW} G.~S. Adkins, C.~R. Nappi and E. Witten, 
\textit{Nucl. Phys.} \textbf{B228}, 552 (1983).

\bibitem{Ko} V.~B. Kopeliovich,
\textit{J. Exp. Theor. Phys.} \textbf{93}, 435 (2001).

\bibitem{MMW} O.~V. Manko, N.~S. Manton and S.~W. Wood,
\textit{Phys. Rev.} \textbf{C76}, 055203 (2007).

\bibitem{BS10} R.~A. Battye and P.~M. Sutcliffe,
\textit{Nucl. Phys.} \textbf{B705}, 384 (2005);
\textit{Phys. Rev.} \textbf{C73}, 055205 (2006).

\bibitem{BMS} R.~A. Battye,  N.~S. Manton and P.~M. Sutcliffe,
\textit{Proc. Roy. Soc.} \textbf{A463}, 261 (2007).

\end{thebibliography}
\end{document}